\documentclass[english]{article}
\usepackage{helvet}
\usepackage[T1]{fontenc}
\usepackage[latin1]{inputenc}
\usepackage{float}
\usepackage{graphicx}

\makeatletter

 \newenvironment{lyxcode}
   {\begin{list}{}{
     \setlength{\rightmargin}{\leftmargin}
     \setlength{\listparindent}{0pt}% needed for AMS classes
     \raggedright
     \setlength{\itemsep}{0pt}
     \setlength{\parsep}{0pt}
     \normalfont\ttfamily}%
    \item[]}
   {\end{list}}

\usepackage{babel}
\makeatother
\begin{document}

\title{Methods for scaling a large member base}

\author{Author: Nathan Boeger (nboeger@khmere.com)}

\maketitle
\begin{abstract}
The technical challenges of scaling websites with large and growing
member bases, like social networking sites, are numerous. One of these
challenges is how to evenly distribute the growing member base across
all available resources. This paper will explore various methods that
address this issue. The techniques used in this paper can be generalized
and applied to various other problems that need to distribute data
evenly amongst a finite amount of resources. 
\end{abstract}

\section{Problem}

Some websites, like social networking sites, are challenged with the
task of trying to scale their exponentially growing member base. How
websites deal with this issue while keeping up with demand varies.
Some websites chose to have a single large database. Then when usage
grows they tend to invest large sums of money on a massive multi-processor
system. Other sites choose to distribute their members across clusters
of inexpensive database servers. This, in theory, will help to scale
the site by adding more database servers when usage increases. However,
the scheme used to segment and distribute the members varies from
site to site. 

One common method used to distribute a sites member's is to segment
the members based on their user name. For example, with a Web site
that contains blogs, 26 different blog databases can be created, based
on alphabetical characters, such as $\{ a,b,c,...,z\}$~all with
identical schematics. Then members can be placed into a database segmented
by the first letter of their user name, allowing for a relatively
intuitive and easy way to distribute the data. Moreover, if the site
requirements include scalability, in theory, 26 different database
servers can be set up, with each server handling its own letter-group.
However, the allocation of the data distribution with such a scheme
will not actually result in even distribution given the actual usage
of letters, and such unevenness will be further augmented when scaled
to a very large member base. To make effective use of resources a
finer method of distributing this data evenly is desirable; this method
should ensure, for example, that each database has as close to the
same number of records as possible.

\subsection{Solution consideration and constraints}

This analysis will find a simple algorithm that will evenly distribute
a large growing population of members across a set of finite servers.
Given the complexity of this problem, measurements for determining
the effectiveness of an algorithm need to be established. One useful
measurement is the Standard Deviation of the distribution%
\footnote{In this paper, the mean of a distribution refers to the ideal mean.
As an example, given 10 buckets and 20 users, the ideal mean is 2
users per bucket.%
}; this measurement describes how evenly spread out the data is%
\footnote{http://en.wikipedia.org/wiki/Standard\_deviation %
}. Thus, the optimal algorithm will produce a small deviation. 

Computational and functional needs of the algorithm also need to be
considered. The optimal algorithm should have the following attributes:

\begin{enumerate}
\item Independence from external data sources - it should require only the
member's user name string and no external data. This is mainly for
simplicity and ease of implementation. 
\item Simple and lightweight - it should not require expensive calculations.
\item Platform agnosticity and independence. 
\item Easily implementable in scripted languages. Scripting languages, such
as PHP and Perl, are very popular and commonly used for Web applications.
Thus, an ideal algorithm would be simple enough to implement within
them. 
\item Must not require the use of large file system subdirectory trees.
Some filesystems, and NAS devices, have limits as to the amount of
sub directories allowed and the algorithm must not exceed them. Also,
for performance reasons, smaller directory trees are desired. 
\item Truly scalable, to very large numbers (at least one billion). 
\item Must be a one-to-one (i.e. injective) function ( for example: $\forall a,b\epsilon f$
we have $f(a)=f(b)\Rightarrow a=b$).
\end{enumerate}

\section{Letter based distributions}

For this paper, I am going to use Fotolog.net as a real world example.
As of writing, Fotolog.net has more than 1 Million registered members.
Each member has the ability to post photos with captions, post entries
into guestbooks, and create lists of friends and favorites. The members
data is spread across several different clusters of database and NAS
servers. Currently, members are distributed based on the first character
of the member's user name. For example, if the member's user name
is \char`\"{}frankie\char`\"{} the member's photo captions are stored
in a table in the 'F' photo database. Because Fotolog.net user names
are comprised of 37 distinct alpha numeric characters%
\footnote{User names are not case sensitive and include the numerals 0 through
9, alphabet characters a through z, and the underscore character (\_). %
}, this creates 37 unique data sources, a data source per every user
name character allowed. 

Similar to the databases, the photos are stored on a NAS file system
and the photos are distributed across the NAS servers based on the
first character of the member's user name. However, unlike the databases,
the photos are kept in a directory structure that can have up to six
sub directories. These sub directories are created based on the expansion
of the characters in the member's users name. For example, the photos
for the member {}``frankie'' would be stored on the NAS server in
the following directory structure:

\begin{lyxcode}
<root~dir>/f/r/a/n/k/i/frankie/my\_photos
\end{lyxcode}
The graph D1%
\footnote{The graph was calculated by mapping the allowed characters (0-9,a-z,\_)
into integers $0-37$ then calculating ~$x\, mod\,37$ where $x$~is
the first character in the user's name. %
} shows Fotolog's first level distribution of member's user names.
Graph D2%
\footnote{The graph was calculated by mapping the allowed characters (0-9,a-z,\_)
into integers $0-37$ then calculating ~$x_{i}\, mod\,37$,$i\epsilon\{0,1\}$
where $i$ specifies the character index in the user's name.%
} shows the distribution of Fotolog's second level letter expansion.
The graph D3%
\footnote{The graph was calculated by mapping the allowed characters (0-9,a-z,\_)
into integers $0-37$ then calculating ~$x_{i}\, mod\,37$,$i\epsilon\{0,1,2\}$
where $i$ specifies the character index in the user's name.%
} shows the distributions of Fotolog's third level expansion. All three
graphs are based on 1,188,968 user names%
\footnote{All current registered members as of April 2, 2005.%
}.

\begin{figure*}
\includegraphics[%
  bb=0bp 0bp 600bp 300bp,
  clip,
  scale=0.6]{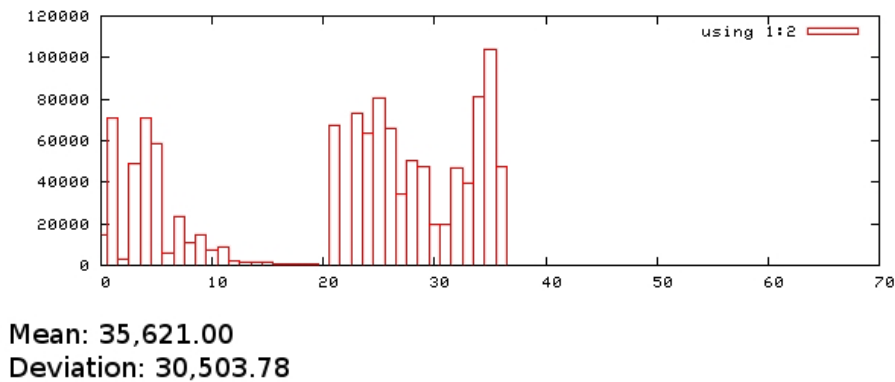}

\caption{Distribution of the first level.}
\end{figure*}

\begin{figure}[H]
\includegraphics[%
  bb=0bp 0bp 600bp 300bp,
  clip,
  scale=0.6]{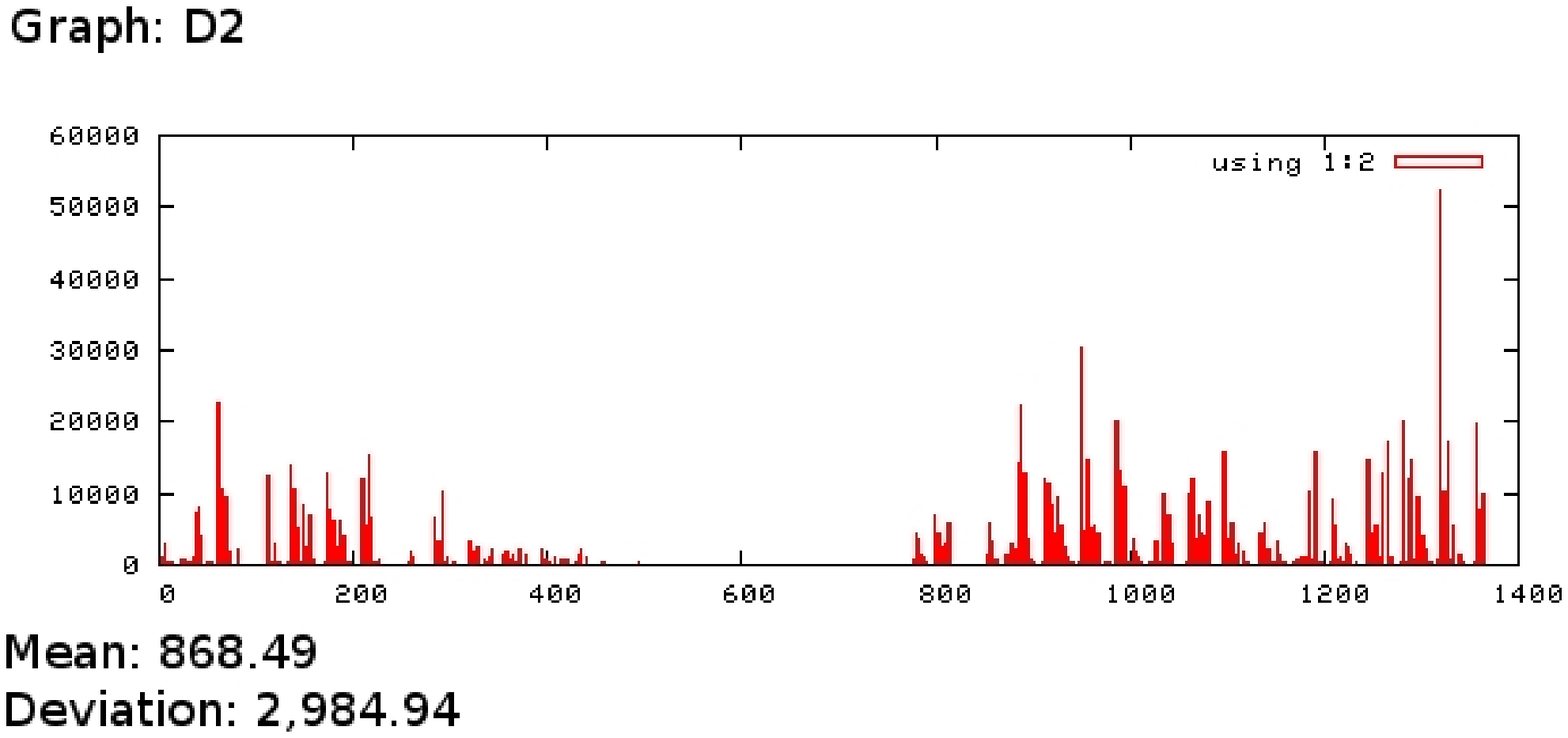}

\caption{Distribution of the second level letter expansion. }
\end{figure}

\begin{figure}[H]
\includegraphics[%
  bb=0bp 0bp 600bp 300bp,
  clip,
  scale=0.6]{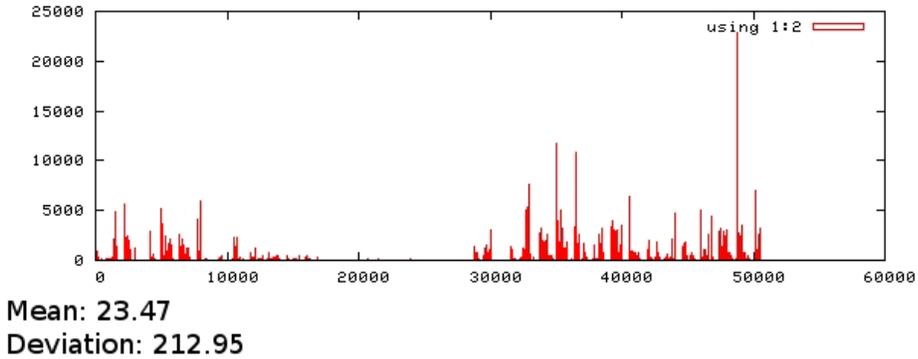}

\caption{Distribution of the third level letter expansion.}
\end{figure}

The distribution shown in graph D1 has a mean of 35,651 with a standard
deviation of 30,504. Because the mean and the standard deviation are
almost equivalent, the letter based distribution algorithm can be
determined to be less than optimal at this level. In fact, if one
were to rearrange the letters one could create a graph that resembles
the Gaussian distribution of random numbers. This would suggest, interestingly
enough, that first letter usage will grow in a Gaussian distribution
like manner. Not very surprising if one considers the members names
as random strings. Then this growth pattern is expected. 

Graph D2 seems to indicate that the distribution will become less
ideal with further letter expansions. The deviation is almost four
times the mean. The graph D3 confirms that skewing of data will become
more extreme the further down the directory structure is traversed.
In this case the mean is nearly ten times smaller than the deviation.
Some buckets directories contain almost 25,000 objects while others
contain almost zero. From these results, I did not see any reason
to continue to create graphs of the distribution further down. It
is obvious from these graphs that the letter based method is not very
effecient. However, it does satisfy many of our constraints. Still,
we should be able to improve on this.

\section{Other possible methods}

An obvious lesson from Fotolog's letter based distribution is that
common letters used in names will tend to skew the distribution. For
example, M and D are very common first letters in names and the graph
D1 clearly shows that these letters do indeed skew the distribution.
Therefore, algorithms such as b-trees, heaps, or any other sorting
algorithms, when based on the characters of the member's user name,
will also result in very uneven distributions, mainly because such
algorithms require subsequent weighting or ordering of data for improved
search results. Not to mention, these algorithms don't make much sense
in this context.

\subsection{ASCII sum algorithm}

A possible algorithm choice is to sum up the ASCII values of the letters
in the member's user name. For example, the user name bob has the
following ASCII values: 98, 111, 98. Its ASCII sum would then be:
98+111+98 = 307. We can then use this sum to distribute the user by
calculating: $x=n\, mod\, S$. Where, n is the ASCII sum, S is a constant
representing the number of resources (also referred to as the bucket
size), and x would then be the data location to store that member.
To calculate the second level data location we would have to modify
the calculation to avoid producing the same number and lumping the
data together into single directory below the top level one %
\footnote{If the calculation $a=x\, mod\,64$ is computed, and then $b=x\, mod\,64$
is computed using the same $x$ value, the result is $a=b$, meaning
that the users will be lumped into a second level directory with the
same number as the top directory. %
}. One simple way we can do this is by removing the first letter in
the member's user name. For example, the member bob would have a second
level calculation by summing the ASCII values of 'o' and 'b'.

Graph N1 was created by calculating $x_{1}=n_{1}\, mod\,31$. Graph
N2 was created by by calculating $x_{2}=n_{2}\, mod\,33$ where $n_{1}$
is the ASCII sum of the entire user name string and $n_{2}$is the
ASCII sum of the member's user name string minus the first character.
The bucket sizes where chosen mostly by experimentation. Prime numbers,
in this case, work most ideally because prime numbers will not have
common factors that could skew the distribution. 

\begin{figure}
\includegraphics[%
  bb=0bp 0bp 600bp 300bp,
  clip,
  scale=0.6]{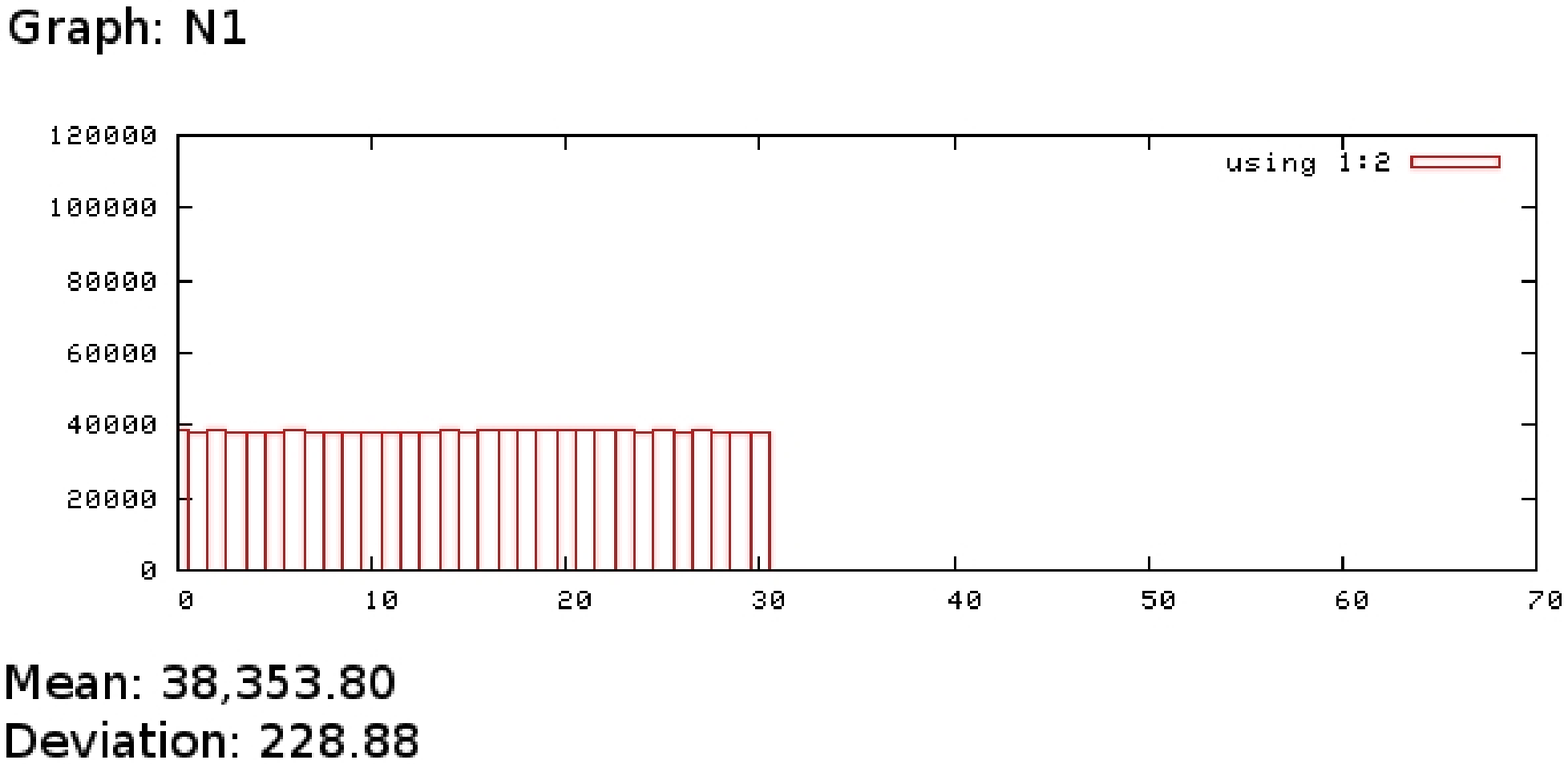}

\caption{Graph N1.}
\end{figure}

\begin{figure}
\includegraphics[%
  bb=0bp 0bp 600bp 300bp,
  clip,
  scale=0.6]{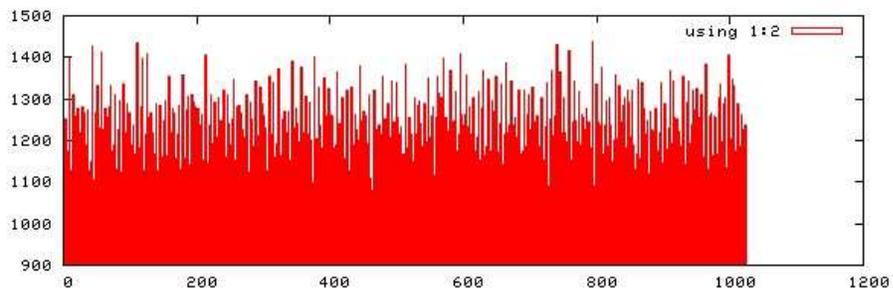}

\caption{Graph N2.}
\end{figure}

The results of Graph N1 appears quite promising; the deviation is
only about 0.006 of the mean. Graph N2 also looks very promising;
the deviation is only about 0.09 of the mean. However, common letter
combinations will slightly skew the distribution. For example, the
names {}``Jonathan, Nathan, Ethan, Nate, Natile'' all have similar
letter combinations. Thus, because addition is commutative, these
names will tend to have similar factors and values when summed. This
might get worse over time, still, this approach seems to be superior
to the letter expansion approach, however, we must continue our search
for a more effecient algorithm.

\subsection{Mapping algorithms}

We could try to create a mapping scheme to scale the member base.
For example, we could keep an increasing counter and for each member
we could assign them the value of the counter when they signed up.
That way we could map a unique integer to a specific user name. We
could then segment the user names based on integer ranges - also referred
to as buckets. For example, members in the integer range$\mathnormal{\left[N_{1},N_{2}\right]}$could
be placed in to a bucket on database $D_{1}$, and so on. Since this
distribution does not consider the members user name string, and since
the number line is smooth, it could possibly provide us with a nice
even distribution. However, the members user name is now dependant
on an external data source. For this method to work, in some form,
their must exist a table that matches a members user name to their
integer I.D. This violates our first constraint and might cause a
slight performance hit when we go to look up the members integer I.D.
Even still, maybe we should toss out the first constraint and keep
this method? It seems to solve our problem. 

I would agree that this approach is attractive at first but it has
some issues. One issue is that we have to be careful not create a
bucket size to large, otherwise, we will have servers with almost
no members on them and others with full buckets. For example, if we
have 24 servers, 1 million members, and if we decide to create a bucket
size of 50,000. This would fill up the first 20 servers while the
other 4 are idle. Now lets say we only have 20 servers and, the same,
1 million members. Once again, if we decide to create a bucket size
of 50,000 all our servers have an even amount of members. However,
where do we place member 1,000,001? At some point in the future, as
the member base grows, we will have a single server with 99,999 members.
While the other 19 have only 50,000. This does seem to be an ideal
situation, but we can make further improvements. 

An obvious approach to improve this situation is to create smaller
buckets and spread these across the servers using a simple calculation.
For example, if we have 20 servers, 1 million members, and a bucket
size of 10,000, we can distribute the resulting 100 buckets by placing
5 onto each server. Then, as our member base grows, at some point
a single server will only have 9,999 more members than the others
\footnote{This will happen when we have 1,049,999 members.%
}. If this is not acceptable, more adjustments can be made to find
the optimal bucket size. The end result is we only need to perform
a simple calculation to find our server like: $x\,=n\, mod\, S$%
\footnote{I have seen another approach in which the servers are mapped onto
the unit circle. The end result is similar since $n\, mod\, S$ is
cyclical. %
}. Where $n$ is the user ID, $S$ is the bucket size, and $x$ is
the server we want. We have a nice bucket size, a fairly even distribution,
and a simple calculation to locate our server. However, we may still
have issues with growth and activity. 

As the member base and level of activity grows, some buckets may have
more activity than others. These buckets with increased activity are
called hot spots. For example, lets say members inside bucket $B_{i}$
become very active and so do the members inside bucket $B_{j}$. Now,
lets say that both of the buckets are stored onto a single server
$S_{k}$ and the server is becoming strained and non-responsive. The
first instinct is to move one of these active buckets off of server
$S_{k}$onto another. However, in doing this you could break your
calculation required to locate the server. Over time these hot spots
could continue to develop and erode away at the bucket size and distribution.
This will eventually force a double look up. The first to find the
users I.D. and the second to locate the bucket and the server it's
located on. The end result is a fragmented and hard to maintain distribution
of buckets. An algorithm for this approach might exist. However, I
cannot seem to come up with one that meets the inital constraints
and is elegant.

\subsection{Hashing algorithms}

One thing we learn from the sum and mapping approaches is that it
would be beneficial if we could find some way to map a member into
the integer number line. This mapping should only use the members
user name, as per our constraints, and still produce some kind of
integer value that does not skew the distribution with common letter
combinations. One such group of algorithms that might solve our problem
are hashing algorithms. One advantageous property of hashing algorithms
is that each unique input should produce a unique output%
\footnote{Hash algorithms do not always produce unique output for every unique
input. However, we will not concern our selves with ways to defeat
hashing algorithms in this paper. %
}%
\footnote{http://en.wikipedia.org/wiki/Hash\_function%
}. This should help us to avoid problems with similar letters used
in member's user names.

Most scripting languages have built in hashing functions, however,
these hashing functions are not standardized. Different scripting
languages could have different hashing implementations. Thus, if we
computed the hash of a string in one language and then we tried to
compute the hash of that same string in another language they might
not match. Also, it is possible that a given language might change
its hashing implementation. Instead we should find a standard and
widely implemented hashing function. 

One excellent candidate is MD5. We could compute the MD5 sum for the
member's user name. Then, the we can distribute the data based on
that result. For example, the name {}``frank'' will have the following
MD5 sum:

\begin{lyxcode}
MD5(frank)~=~d268c8fe7f154537c2c9ed60a0b8f2fd
\end{lyxcode}
We could chose the first character from the MD5 sum (in this case
{}``d''), and then distribute the data based on this character.
However, MD5 returns its results in ASCII hexadecimal format. Thus,
the first character will only have 16 different values, forcing the
top level bucket size to be too small because the limit can be extended
to only 16 different data sources. Using the first two characters
of the MD5 sum may appear to be a solution, but that will yield 255
top level buckets, which might be excessive because some of these
data sources could be file system mounts%
\footnote{This is a purely arbitrary restriction; as a Unix administrator keeping
track of more than 254 mount points on every server is simply not
an appealing option. %
}. 

Another approach is to take these first two characters and mod them
by 64. This would create a total of 64 data sources, which is easy
to maintain and, if the data is spread out evenly, 1 billion objects
would only create about 16 million entries per data source $(\frac{2^{30}}{2^{6}}=2^{24})$.
The graph M1 demonstrates the distribution of this algorithm.

\begin{figure}[H]
\includegraphics[%
  bb=0bp 0bp 600bp 300bp,
  clip,
  scale=0.6]{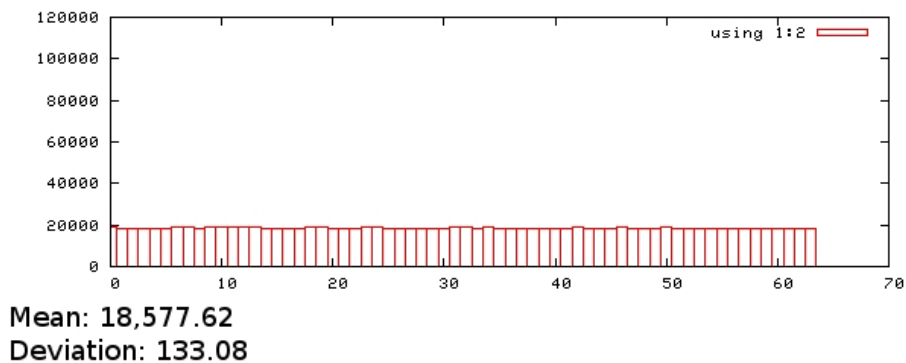}

\caption{Graph of the first level of the MD5 distribution.}
\end{figure}

As seen from the graph M1, the MD5 algorithm provides a nice and even
distribution. The average bucket holds around 18,000 objects and the
deviation is about 0.007 of the mean. This algorithm appears to perform
ideally for the top level directories and database distribution. However,
consider the subdirectories. 

When calculating the value for the second level, the currently calculated
value for the top level directory must not be reused; otherwise, all
the data will end up lumped together in a single directory below the
top level directory. The second level is calculated using the next
two digits from the MD5 string (places 2 and 3) for the mod 64 calculation.
Graph M2 shows this data.

\begin{figure}[H]
\includegraphics[%
  bb=0bp 0bp 600bp 300bp,
  clip,
  scale=0.6]{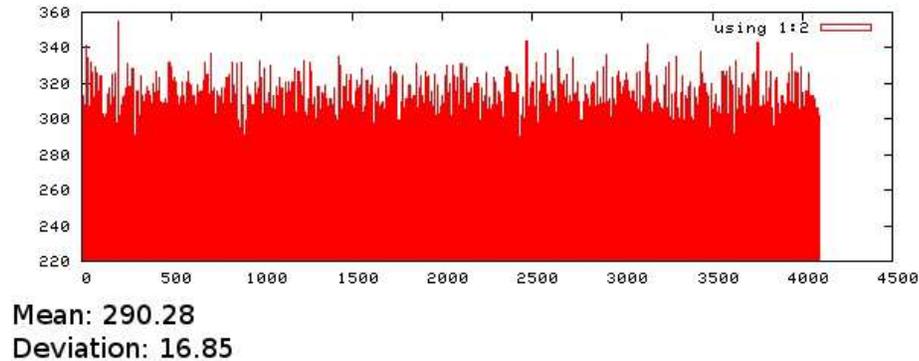}

\caption{Graph of the second level of the MD5 distribution. }
\end{figure}

The graph M2 shows that although the results are not as optimal as
the top level directory, the deviation is only about 0.06 of the mean.
Each bucket has about 300 objects and could easily handle many more
without going beyond file system limits%
\footnote{File system limits depend on the filesystem; it is assumed that the
readers are using fairly modern file systems.%
}. The same approach can also be used for the final directory level;
the next two digits, in places 4 and 5, from MD5 are be used for the
final calculation. However, instead of limiting this level to 64 it
should be increased to 128 in order to increase the total amount of
buckets and decrease the total amount of data per bucket. Also, note
that from a given top level directory, only a total of: $65\times128=8192$
subdirectories are created, which is well below most file system limitations%
\footnote{Again, this depends on the particular file system. Some major NAS
devices have limits of 64,000 files per directory; make sure to be
well below this limit.%
}. Graph M3 shows the distribution at the third and final level.

\begin{figure}[H]
\includegraphics[%
  bb=0bp 0bp 600bp 300bp,
  clip,
  scale=0.6]{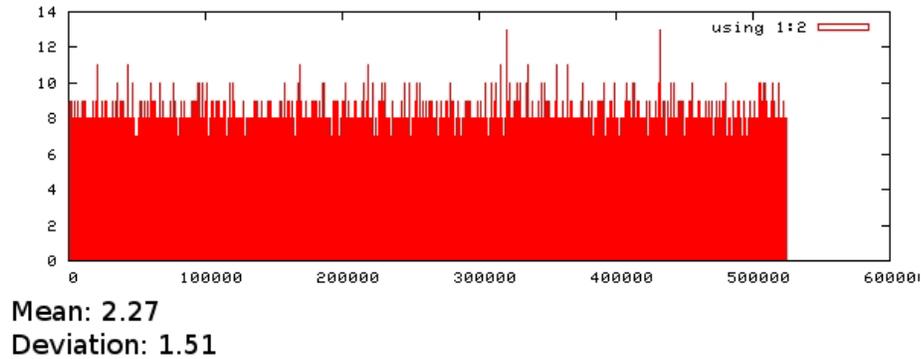}

\caption{Graph of the third level of the MD5 distribution. }
\end{figure}

As the graph M3 shows, the deviation is almost half of the mean. Normally
this would not be ideal. However, at this level, the current data
set is too small to fully fill all: $64\times64\times128=524,288$
buckets. Regardless, even if the mean remained constant and the object
inventory grew one thousands times, each bucket would only contain
2,300 objects%
\footnote{Note: $2.27\times1000\approx2,300$%
}. This is much lower than the letter based scheme in which some buckets
contain more than 25,000 objects.

It would seem that the distribution using MD5 is far superior to the
other approachs we have examined. Also, the MD5 algorithm will provide
us with enough buckets to avoid having to descend down further than
three levels.

\noindent The MD5 algorithm used to calculate a member's top level
location would be:

\begin{enumerate}
\item MD5 the member's user name 
\item Convert the first two characters, places 0 and 1, from the MD5 into
an integer: $x$
\item Calculate: $a=x\, mod\,64$. then use $a$ and the member's users
name to locate the users data
\end{enumerate}
The MD5 algorithm used to calculate a member's third level location
would be:

\begin{enumerate}
\item MD5 the member's user name 
\item Convert the first two characters, places 0 and 1, from the MD5 into
an integer: $x$
\item Convert characters in places 2 and 3 from the MD5 into an integer:
$y$
\item Convert characters in places 4 and 5 from the MD5 into an integer:
$z$
\item Calculate: $a=x\, mod\,64$, $b=y\, mod\,64$, $c=z\, mod\,64$
\item The full data path would then be: <NAS mount>/a/b/c/<user\_name>
\end{enumerate}

\section{Conclusion}

In this paper we have examined several methods used to distribute
a large growing member base across a finite amount of resources. From
our data we have observed that the best algorithm to yield a fairly
even distribution is to use the results of the MD5 hash of the member's
user name. From the top level down to the third level the distribution
will remain fairly even. Also, the MD5 distribution satisfies all
initial constraints.

\begin{enumerate}
\item The algorithm does not require any external data sources other than
the user name.
\item Processing MD5 strings is not very expensive for small strings such
as user names%
\footnote{R. Rivest, rfc1321, 1992.%
}.
\item The algorigthm in general is platform agnostic.
\item Most popular scripting languages have built in support for MD5%
\footnote{As of writing: PHP, Perl, Tcl, and Python support MD5. Others can
be found at: http://userpages.umbc.edu/\textasciitilde{}mabzug1/cs/md5/md5.html%
}. Therefore, its very easy to implement within a scripting language.
\item For Fotologs implementation, the MD5 algorithm will provide enough
buckets to avoid having to descend down further than three levels. 
\item We have seen from the graphs that this will scale well beyond 1 billion
members.
\item It is a one to one function%
\footnote{MD5 was designed to be a message-digest algorithm and is known to
have collision weaknesses. These can be avoided by using the full
user name as the final data location (ex: /a/b/c/<user\_name>) and
therefore claim this mapping to be one to one.%
}. 
\end{enumerate}
I have to admit that using MD5, or hashing functions in general, to
distribute data is not unique. And I am not trying to take credit
for it. After doing my research and writing this paper, I have seen
what appear to be similar approaches. However, I have not found any
source that documents this approach. Therefore, I still felt that
this paper was worth writing and publishing even if its only for documentation
purposes.~\\
\\
Author email: nboeger@khmere.com ~\\

\$RCSfile: md5\_algo.lyx,v \$,\$Revision: 1.13 \$
\end{document}